\documentclass[sn-aps]{sn-jnl}

\jyear{2021}%

\theoremstyle{thmstyleone}%
\theoremstyle{thmstyletwo}%

\theoremstyle{thmstylethree}%

\begin{document}

\title[Article Title]{Vinen's latest thoughts on the ``bump'' puzzle in decaying He II counterflow turbulence}

\author[1,2]{\fnm{Wei} \sur{Guo}}
\email{wguo@magnet.fsu.edu}
\author[1,3]{\fnm{Toshiaki} \sur{Kanai}}

\affil[1]{\orgdiv{National High Magnetic Field Laboratory}, \orgaddress{\street{1800 East Paul Dirac Drive}, \city{Tallahassee}, \postcode{32310}, \state{Florida}, \country{U.S.A}}}

\affil[2]{\orgdiv{Mechanical Engineering Department}, \orgname{FAMU-FSU College of Engineering, Florida State University}, \orgaddress{\city{Tallahassee}, \postcode{32310}, \state{Florida}, \country{U.S.A}}}

\affil[3]{\orgdiv{Department of Physics}, \orgname{Florida State University}, \orgaddress{\city{Tallahassee}, \postcode{32306}, \state{Florida}, \country{U.S.A}}}

\abstract{The pioneering work of William F. Vinen (also known as Joe Vinen) on thermal counterflow turbulence in superfluid helium-4 largely inaugurated the research on quantum turbulence. But despite decades of research on this topic, there are still open questions remaining to be solved. One such question is related to the anomalous increase of the vortex-line density $L(t)$ during the decay of counterflow turbulence, which is often termed as the ``bump'' on the $L(t)$ curve. In 2016, Vinen and colleagues developed a theoretical model to explain this puzzling phenomenon (JETP Letters, \textbf{103}, 648-652 (2016)). However, he realized in the last a few years of his life that this theory must be at least inadequate. In remembrance of Joe, we discuss in this paper his latest thoughts on counterflow turbulence and its decay. We also briefly outline our recent experimental and numerical work on this topic.}



\maketitle

\section{Introduction}\label{sec1}
When liquid $^4$He is cooled to below about 2.17~K, it enters the superfluid phase (known as He II)~\cite{Tilley-1990-book}. Phenomenologically, He II can be considered as a mixture of two miscible fluid components: an inviscid superfluid and a viscous normal fluid that consists of thermal quasiparticles (i.e., phonons and rotons)~\cite{Landau-book}. The flow of the superfluid is irrotational, and any rotational motion in a simply-connected volume can emerge only with the formation of topological defects in the form of quantized vortex lines. These vortex lines are density-depleted thin tubes, each carrying a quantized circulation of $\kappa=h/m$, where $h$ is Planck's constant and $m$ is the mass of a helium atom~\cite{Donnelly-1991-B}. As a two-fluid system, He II has many unique thermal and mechanical properties. For instance, bulk He II can support two sound-wave modes: an ordinary pressure-density wave (i.e., the first sound) where the two fluids oscillate in phase, and a temperature-entropy wave (i.e., the second sound) where the two fluids oscillate oppositely~\cite{Landau-book}. Furthermore, heat transfer in He II is via a counterflow mode~\cite{Landau-book}: the normal fluid moves in the direction of the heat flux $q$ with a mean velocity $U_n=q/\rho{s}T$, where $\rho=\rho_s+\rho_n$ is the total density of He II, $s$ is its specific entropy, and $T$ is the temperature; The superfluid moves in the opposite direction with a mean velocity $U_s=(\rho_n/\rho_s)U_n$ to ensure no net mass flow.

In the pioneering work of Joe Vinen, he discovered that turbulence can spontaneously emerge in counterflow in a uniform channel when the relative velocity $U_{ns}=\lvert U_n- U_s\rvert$ exceeds a small critical value $U_c$~\cite{Vinen-1957-PRS-I,Vinen-1957-PRS-II,Vinen-1957-PRS-III,Vinen-1957-PRS-IV}. A phenomenological theory was proposed by him at the same time~\cite{Vinen-1957-PRS-III}, and a more detailed understanding was achieved later by Schwarz who developed a vortex filament model to simulate the counterflow turbulence~\cite{Schwarz-1977-PRL,Schwarz-1988-PRB}. More sophisticated simulations were reported subsequently by other researchers~\cite{Adachi-PRB-2010,Kivotides-PRB-2008,Baggaley-PRE-2014}. According to these theoretical works, the turbulence exists only in the superfluid and is induced by a more or less random tangle of quantized vortices. A mutual friction force between the two fluids then emerges due to the scattering of the thermal quasiparticles off the quantized vortices~\cite{Vinen-1957-PRS-III,Schwarz-1988-PRB}. This model also nicely explains the observed $U_{ns}$ dependance of the vortex-line density $L$ in steady-state counterflow, i.e., $L^{1/2}=\gamma (U_{ns}-U_c)$ with $\gamma$ being a temperature-dependant coefficient.

However, extensive experimental studies by Tough and colleagues indicated that counterflow turbulence may be more complex~\cite{Tough-1982-PLTP}. They demonstrated that there can be two turbulent regimes in a uniform channel with relatively small cross-sectional area: a T-I state with smaller values of $\gamma$ and a T-II state with larger values of $\gamma$. They proposed that transitions to turbulence in the normal fluid may be responsible. In larger channels, they found a transition from laminar flow directly to a turbulent state denoted as T-III, and they suggested that both fluids might be turbulent in T-III. Later, Melotte and Barenghi developed a theory showing that the T-I to T-II transition may be associated with an instability in the laminar flow of the normal fluid~\cite{Melotte-1998-PRL}. Possible existence of normal-fluid turbulence in counterflow was indeed indicated in some early experimental studies~\cite{Allen-PRS-1965,Tough-PR-1965}. More specific evidence showing the laminar-to-turbulent transition of the normal fluid in counterflow was provided later by Guo \textit{et al.}, who used metastable He$^*_{2}$ molecular tracers to visualize the normal fluid flow~~\cite{Guo-2010-PRL}. Since then, there have been various measurements of the normal fluid velocity field in steady counterflow, revealing the presence of large-scale turbulence in the normal fluid with non-classical second-order statistics~\cite{Chagovets-PF-2011,LaMantia-JFM-2013,Kubo-JLTP-2019,Marakov-2015-PRB,Gao-2017-PRB,Gao-2017-JLTP,Bao-2018-PRB,Gao-2018-PRB}.

Besides the studies on steady-state counterflow, many experiments have also been reported on the decay of counterflow turbulence when the heat flux is turned off~\cite{Vinen-1957-PRS-II,Skrbek-2003-PRE,Gordeev-JLTP-2005,Gao-2016-PRB,Gao-2016-JETP,Babuin-2016-PRB}. According to Vinen's model, the decay of the vortex-line density is given by:
\begin{equation}
\frac{dL}{dt}=-\frac{\chi_2 \kappa}{2\pi} L^2,
\label{eq1}
\end{equation}
where $\chi_2$ is a dimensionless parameter of order unity. Therefore, one would expect to see a monotonic decay of the line density as $L(t)\propto (t+t_0)^{-1}$. However, in the earliest work by Vinen~\cite{Vinen-1957-PRS-II}, it was noted that $L$ did not always decay monotonically, i.e., a ``bump" can appear at about one second after the heater was turned off. Skrbek \textit{et al.} first realized that the line density in the final stage of the decay actually scaled as $L(t)\propto t^{-3/2}$~\cite{Skrbek-2003-PRE}, which is consistent with the decay of a quasiclassical turbulence having a Kolmogorov energy spectrum~\cite{Stalp-1999-PRL}. Such a quasiclassical turbulence can emerge in He II when the two fluids are strongly coupled by mutual friction at length scales greater than the mean vortex-line spacing $\ell=L^{-1/2}$. However, the mechanism underlying the observed bump remained a mystery for many years. Several theories have been proposed to explain the origin of the bump~\cite{Schwarz-PRL-1991,Barenghi-2006-PRE,Mineda-2013-JLTP}, but a complete understanding of the phenomenon requires experimental information on the velocity field in counterflow besides just the time variation of $L$. This information was provided by Gao \emph{et al.} in a more recent experiment where both quantitative flow visualization and second-sound attenuation measurements were incorporated~\cite{Gao-2016-PRB}. These authors reported that the anomalous decay of $L(t)$ was always correlated with large-scale normal-fluid turbulence in steady counterflow before the heater was switched off. This observation inspired Vinen to develop a theoretical model of the bump. At the same time, a similar theory was developed independently by L'vov and Pomyalov. Together with the experimental teams, they published a few joint papers in 2016 to report this progress~\cite{Gao-2016-JETP,Babuin-2016-PRB}. The essence of the bump theory was later adapted by Walmsley and Golov to explain similar bumps observed in the decay of superfluid turbulence in the $T=0$ limit~\cite{Walmsley-2017-PRL}.

During a visit to our group in 2019, Vinen realized that a key assumption in his model was likely incorrect. Since then, he focused on this problem until he passed away in 2022. In remembrance of Joe Vinen, we discuss his latest thoughts in this paper. In Sec.~\ref{Sec2}, we review the bump theory and some key observations in our experiments. In Sec.~\ref{Sec3}, we discuss Vinen's thoughts on why the model needs to be revised. Sec.~\ref{Sec4} summarizes the content of this paper and also briefly outlines our ongoing experimental and numerical work on counterflow turbulence in He II.

\section{Variation of $L(t)$ in decaying counterflow}\label{Sec2}
\begin{figure*}[t]
\centering
\includegraphics[width=1\linewidth]{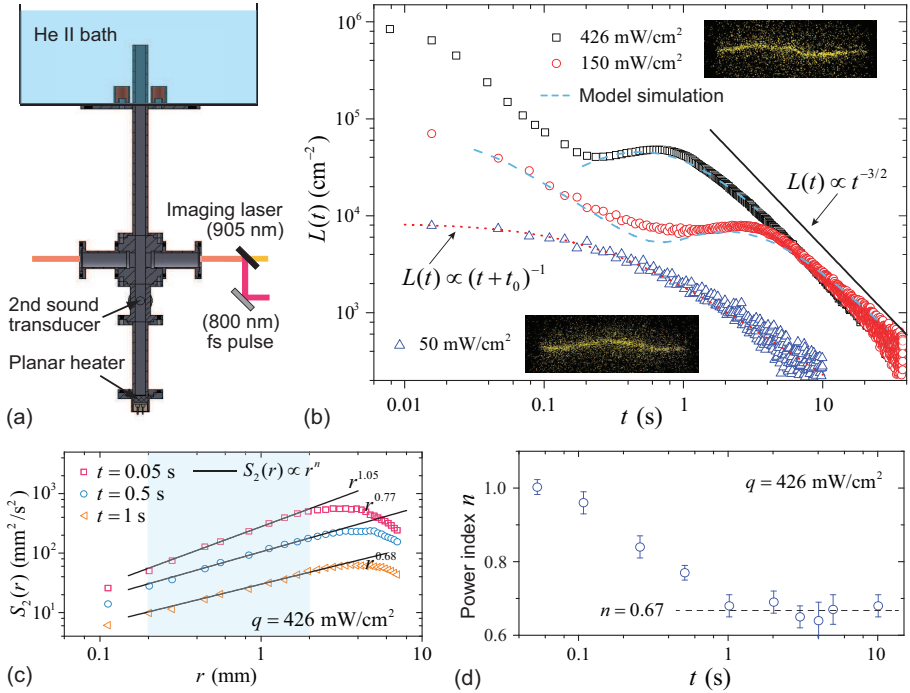}
\caption{(a) A schematic of the He II counterflow experimental setup. (b) Decay of the vortex-line density $L(t)$. The measurements were conducted at 1.65 K, and $t=0$ denotes the moment when the heater was turned off. The dashed blue curves are model simulations as reported in Ref.~\cite{Gao-2016-JETP}. The inset images show the deformation of representative He$^*_2$ molecular tracer lines created in steady counterflow before the heater was turned off. (c) Measured second-order transverse velocity structure function of the normal fluid $S_2(r)$ at different decay times. The solid lines represent power-law fits to the data in the shaded region. (d) The power-law index extracted from (c) as a function of the decay time $t$.
}
\label{Fig1}
\end{figure*}

In the experiment reported by Gao \emph{et al.}~\cite{Gao-2016-PRB}, a vertical flow channel with a square cross-section (side width: 9.5 mm; length 300 mm) was connected to a temperature-controlled He II bath (see the setup schematic in Fig.~\ref{Fig1}~(a)). A planar heater was installed at the bottom of the flow channel to generate thermal counterflow. To probe the normal-fluid motion, a femtosecond laser pulse (wavelength: 800~nm; pulse length: about 30 fs)~\cite{Gao-2015-RSI} was focused to pass through the channel to created a thin horizontal line of He$^*_2$ excimer molecules with a thickness of about 100~$\mu$m and a length of about 1~cm. These molecular tracers, which are entrained by the viscous normal fluid, can be imaged via laser-induced fluorescence driven by an imaging laser pulse at 905~nm~\cite{Guo-2010-PRL}. By examining the displacement of tracer-line segments, one can determine the local normal-fluid velocity in the heat flux direction $U^{(z)}_n(r)$ at any location $r$ along the tracer line. At the meanwhile, standard second-sound attenuation measurements were conducted to determine the spatially averaged vortex-line density $L(t)$ in the channel using a pair of porus membrane-based second-sound transducers~\cite{Mastracci-2018-RSI}.

The key observations are summarized in Fig.~\ref{Fig1}~(b)-(d). At relatively large heat fluxes in steady counterflow, we found that as the heater was turned off, $L(t)$ first dropped drastically and then a bump emerged. At large decay times, $L(t)$ appeared to decay as $L(t)\propto t^{-3/2}$. These results are in good agreement with the earlier observations of Skrbek \textit{et al.}~\cite{Skrbek-2003-PRE}. However, at sufficiently low heat fluxes, we discovered that $L(t)$ indeed decayed as $L(t)\propto(t+t_0)^{-1}$, which is what one would expect for the decay of a random tangle of vortices according to Eq.~\ref{eq1}. More importantly, our flow visualization measurements revealed that the appearance of the bump was correlated with the large-scale turbulence in the steady state. At large heat fluxes where the bump was seen, the normal-fluid flow in the steady counterflow was always turbulent since we observed random deformations on the tracer lines (see inset in Fig.~\ref{Fig1}~(b)). On the other hand, at small heat fluxes where the bump was absent, the tracer lines deformed in a reproducible manner in the steady counterflow, suggesting laminar normal-fluid flow.

The deformation of the tracer lines also allowed us to evaluate the second-order transverse velocity structure function defined as $S_2(r)=\langle(U^{(z)}_n(r+r_0)-U^{(z)}_n(r_0)^2\rangle$, where the angled brackets denote an ensemble average over various reference location $r_0$ and different experimental runs~\cite{Gao-2016-PRB}. The obtained $S_2(r)$ at different decay times with a steady-state heat flux of $q=426$~W/cm$^2$ is shown in Fig.~\ref{Fig1}~(c). In the shaded region, the $S_2(r)$ data can be fitted with a power-law function $S_2(r)\propto r^n$. The extracted power index $n$ as a function of the decay time $t$ is shown in Fig.~\ref{Fig1}~(d). It is clear that $n$ deceases with $t$ from about 1 in steady state to about 2/3 at $t\simeq1$~s, i.e., the time around which the bump of $L(t)$ appears. Beyond this time, $n$ settles at 2/3. Note that the second-order velocity structure function and the turbulence energy spectrum are connected through a bridge relation as detailed in Ref.~\cite{Tang-2021-PRB}. Within the scaling length-scale range, the power-law form of $S_2(r)\propto r^n$ corresponds to a scaling of the energy spectrum $E(k)\sim k^{-(n+1)}$. Our data suggests that in decaying counterflow with a high steady-state heat flux, the energy spectrum evolves from an approximate form of $E(k)\sim k^{-2}$ to the classical Kolmogorov form of $E(k)\sim k^{-5/3}$ as the turbulence decays, and the bump of $L(t)$ appears upon the completion of the spectrum evolution. We would also like to point out that we calculated the energy spectrum in steady counterflow directly using our flow visualization data, as documented in Ref.~\cite{Bao-2018-PRB}. Our analysis confirmed that the energy spectrum in steady counterflow exhibits a power-law scaling with a power index of about 2, consistent with the scaling of the structure function.

Inspired by these observations, Joe proposed an appealing explanation of the bump. Note that as the heat current was switched off, the two fluids can become strongly coupled by the mutual friction in a few milliseconds~\cite{Vinen-2000-PRB}. Assuming homogeneous and isotropic flows, the total energy $\mathscr{E}$ per unit He II mass is approximately given by $\mathscr{E}=\mathscr{E}_1+\mathscr{E}_2$, where $\mathscr{E}_1=B(\rho_s/\rho){\kappa}^{2}L$ accounts for the flows associated with individual vortices at scales comparable or smaller than $\ell$ (here $B$ is a dimensionless factor of order unity)~\cite{Donnelly-1991-B}, and $\mathscr{E}_2$ represents the kinetic energy density associated with the large-scale coupled flows. The decay rate of the turbulence energy is related to the vortex-line density $L$ as~\cite{Stalp-1999-PRL}:
\begin{equation}
\frac{d\mathscr{E}}{dt}=B{\kappa}^{2}\frac{\rho_s}{\rho}\frac{dL}{dt}+\frac{d\mathscr{E}_2}{dt}=-{\nu'}({\kappa}L)^2,
\label{eq2}
\end{equation}
where $\nu'$ is an effective viscosity of He II~\cite{Stalp-2002-PF,Chagovets-2007-PRE,Gao-2016-PRB}. If there are no large-scale flows, i.e., $\mathscr{E}_2=0$, Eq.~(\ref{eq2}) reduces to exactly Eq.~(\ref{eq1}), which essentially describes the decay of a random tangle of vortices. But when $\mathscr{E}_2$ is nonzero, i.e., there are large-scale flows induced by polarized vortex bundles, the variation of $L(t)$ can deviate from the $(t+t_0)^{-1}$ scaling.

For instance, if the large-scale flows exhibit a Kolmogorov spectrum so that the energy cascade rate $\epsilon(k)$ remains constant in the inertial $k$ range, $\dot{\mathscr{E}}_2$ can be estimated based on the energy cascade rate at the integral scale, i.e., $\dot{\mathscr{E}}_2=-\epsilon(k_D)\simeq-\Delta U_D^2/2\tau_D$, where $\Delta U_D$ denotes the velocity variance at the integral scale (which is comparable to the channel width $D$) and $\tau_D\simeq D/\Delta U_D$ is the turnover time of the large-scale eddies. Then, considering the fact that $\Delta U_D$ decreases with time as the turbulence decays~\cite{Stalp-1999-PRL}, one can derive $\dot{\mathscr{E}}_2(t)\simeq-\Delta U_D(0)^3/2D[1+t/{\tau}_D(0)]^3$~\cite{Gao-2016-JETP}. Adding this $\dot{\mathscr{E}}_2$ to Eq.~(\ref{eq2}) can lead to a smooth transition from the scaling of $L(t)\propto(t+t_0)^{-1}$ at small $t$ to the scaling of $L(t)\propto t^{-3/2}$ at large $t$.

However, the situation changes when the large-scale flows have an initial spectrum steeper than the Kolmogorov form, i.e., $E(k)\sim k^{-2}$. In this case, it is straightforward to derive that the energy cascade rate depends on $k$ as $\epsilon(k)\sim k^{-1/2}$. Therefore, initially $\dot{\mathscr{E}}_2\simeq-\epsilon(k_{\ell})$, which is smaller than $-\epsilon(k_D)$ by a factor $(k_{\ell}/k_D)^{-1/2}$ and hence should be negligible under typical experimental conditions. Only after the spectrum evolves into the Kolmogorov form, $\dot{\mathscr{E}_2}$ can rise to $-\epsilon(k_D)$ and contribute to the buildup of the vortices. This delayed cascade of the large-scale turbulence energy can give rise to a bump of $L(t)$. Based on this idea, Joe proposed that $\dot{\mathscr{E}}_2$ in decaying counterflow could be modeled as:
\begin{equation}
\dot{\mathscr{E}}_2\simeq -\frac{\Delta U(0)^3}{2D[1+t/\tau_D(0)]^3}F(t),
\label{eq3}
\end{equation}
with $F(t)$ being a dimensionless function that evolves smoothly from 0 to 1 over a time comparable to $\tau_D(0)\simeq D/\Delta U_D(0)$~\cite{Gao-2016-JETP}. It turns out that numerical simulations based on Eqs.~(\ref{eq2}) and (\ref{eq3}) can reasonably reproduce the location and the height of the bump, as shown in Fig.~\ref{Fig1}~(b).

\section{Vinen's latest thoughts}\label{Sec3}
A key assumption made in the theory presented in Sec.~\ref{Sec2} is that the time it takes for the initial energy spectrum (i.e., $E(k)\sim k^{-2}$) to evolve to the final Kolmogorov form is about the turnover time of the large-scale eddies, i.e., $\tau_D(0)\simeq D/\Delta U_D(0)$. But is this assumption valid? During a visit to our lab in 2019, Joe raised this question to us. Based on his suggestion, we adopted the Leith diffusion model to examine the time evolution of $E(k,t)$~\cite{Leith-1967-PF}:
\begin{equation}
\frac{\partial E(k,t)}{\partial t}=\frac{2}{11C^{3/2}}\frac{\partial}{\partial k}\left[k^{13/2}\frac{\partial}{\partial k}(k^{-3}E^{3/2}(k))\right]-2\nu k^2E(k),
\label{eq4}
\end{equation}
where $C=1.71$ is the Kolmogorov constant and $\nu$ is the kinematic viscosity of He II. This equation applies to homogeneous and isotropic turbulence and has been widely utilized to study energy-spectrum evolution in classical fluids~\cite{Grebenev-JP-2014, Nazarenko-JP-2017, Rubinstein-CF-2017}. In decaying counterflow where the two fluids are strongly coupled, one would expect that the spectrum evolution of the large-scale flows may also obey this classical model.

\begin{figure*}[t]
\centering
\includegraphics[width=1\linewidth]{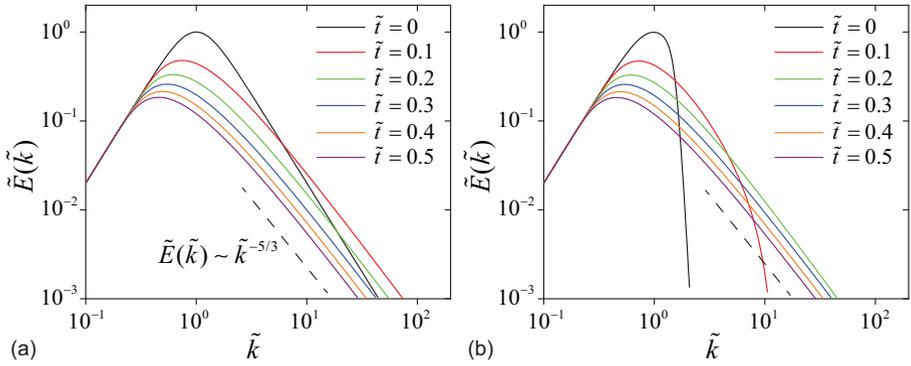}
\caption{Evolution of the dimensionless energy spectrum with an initial form given by (a) Eq.~(\ref{eq6}) and (b) Eq.~(\ref{eq7}). These simulations were conducted using the Leith diffusion equation as discussed in the text. The dashed line shows the scaling of a Kolmogorov spectrum.
}
\label{Fig2}
\end{figure*}

In our study, we suppose that initially the energy spectrum $E(k,0)$ is proportional to $k^2$ at small $k$, which reaches a maximum $E_0$ at $k=k_0$, and then falls off as $k^{-2}$ at large $k$ as indicated by our flow visualization data. We take the turnover time for the energy-containing eddies as $\tau_0=2^{3/2}\pi/k^{3/2}_0E^{1/2}_0$. This is equivalent to taking $k_0=2\pi/D$ and $\tau_0=2D/\Delta U_0$. We can then introduce the following dimensionless parameters $\tilde{E}=E/E_0$, $\tilde{t}=t/\tau_0$, and $\tilde{k}=k/k_0$ to convert Eq.~(\ref{eq4}) into a dimensionless form:
\begin{equation}
\frac{\partial \tilde{E}}{\partial \tilde{t}}=\frac{\partial}{\partial \tilde{k}}\left[\tilde{k}^{13/2}\frac{\partial}{\partial \tilde{k}}(\tilde{k}^{-3}\tilde{E}^{3/2})\right].
\label{eq5}
\end{equation}
Here we have dropped the viscous term in Eq.~(\ref{eq4}) and replaced it by imposing a sharp cutoff at $\tilde{k}=1000$. Our initial energy spectrum then takes the following dimensionless form (i.e., see the black line in Fig.~\ref{Fig2}~(a)):
\begin{equation}
\tilde{E}(\tilde{k},0)=\frac{2\tilde{k}^2}{1+\tilde{k}^4}.
\label{eq6}
\end{equation}
By integrating Eq.~(\ref{eq5}), the evolution of the spectrum can be obtained, which is shown in Fig.~\ref{Fig2}~(a). We see that the spectrum evolves to the Kolmogorov form after a dimensionless time $\tilde{t}$ less than 0.1, i.e., about an order of magnitude smaller than the turnover time of the energy-containing eddies. Indeed, this is not really a surprising result because only a small fraction of the energy in the energy-containing eddies needs to be lost in order for the evolution to the Kolmogorov spectrum to occur at large $k$.

We have also tested the evolution of an initial energy spectrum that has practically no weight for $\tilde{k}$ greater than about 2:
\begin{equation}
\tilde{E}(\tilde{k},0)=\frac{\tilde{k}^2}{1+\tilde{k}^4}\left[1-\tanh(5(\tilde{k}-1.5))\right].
\label{eq7}
\end{equation}
The simulation result is shown in Fig.~\ref{Fig2}~(b). Although this spectrum evolves into the Kolmogorov form more slowly than does that given by Eq.~(\ref{eq6}), the time required for this evolution is still significantly less than the turnover time of the energy containing eddies. Therefore, the analysis based on the Leith equation clearly suggests that the evolution of the initial energy spectrum to the Kolmogorov form in decaying counterflow should be much quicker than what we had believed, i.e., the $F(t)$ function in Eq.~(\ref{eq3}) should evolve from 0 to 1 in a much shorter time than $\tau_D(0)$. According to the theory presented in Sec.~\ref{Sec2}, the bump should therefore emerge much earlier than the observed time. On the other hand, experimental data shown in Fig.~\ref{Fig1}~(d) do suggest that the spectrum evolves to the Kolmogorov form in a time comparable to $\tau_D(0)$. This discrepancy made Joe realize that some ingredients may be missing in his original theoretical model.

\begin{figure*}[t]
\centering
\includegraphics[width=1.0\linewidth]{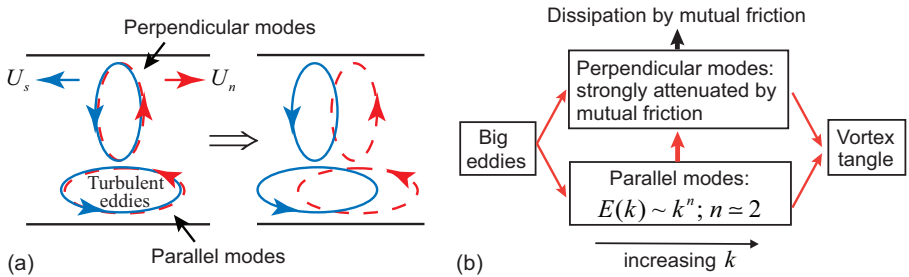}
\caption{(a) A schematic showing how the turbulent eddies in the two fluids of different orientations become decoupled in He II counterflow. See more detailed discussions in Refs.~\cite{Biferale-2019-PRL,Biferale-2019-PRB}. (b) A schematic illustrating Joe Vinen's view on how the turbulence energy flows in counterflow.
}
\label{Fig3}
\end{figure*}

Note that the Leith equation (i.e., Eq.~\ref{eq4}) applies to isotropic turbulence. But as Biferale \emph{et al.} pointed out recently, steady-state counterflow turbulence can exhibit strong anomalous anisotropy at small scales~\cite{Biferale-2019-PRL,Biferale-2019-PRB}. This anisotropy can be conceptually understood by considering two turbulent eddies (one in each fluid) as shown schematically in Fig.~\ref{Fig3}~(a). If these two eddies do not correlate and overlap in space, they would be very effectively damped by the mutual friction~\cite{Vinen-2000-PRB}. On the other hand, if the two eddies are coupled, the mutual friction dissipation may remain small such that these eddies could survive for a sufficient time to sustain a cascade. But due to the opposite mean flows, any initially coupled eddies must be swept apart at later times. The larger eddies can remain coupled for longer times, whereas the smaller eddies become uncorrelated quickly and hence are promptly damped. Therefore, the population of the coupled eddies must be suppressed as the length scale is reduced. Furthermore, for coupled eddies that are elongated perpendicular to the heat flux direction (i.e., Fourier modes with large $k_z$, which are denoted by Joe as the ``perpendicular modes''), they remain coupled for much shorter times as compared to those eddies elongated parallel to the heat flux direction (i.e., Fourier modes with small $k_z$, denoted as the ``parallel modes''). Therefore, the perpendicular modes should be strongly suppressed in steady counterflow.

Joe believed that this anisotropy in steady counterflow must affect how fast the energy spectrum could evolve into the Kolmogorov form after the heat flux was switched off. His idea is illustrated in Fig.~\ref{Fig3}~(b). In the steady state, the large-scale eddies may feed energy to both the perpendicular modes and the parallel modes. The perpendicular modes are strongly damped by the mutual friction. Consequently, the parallel modes must lose energy by inertial transfer to both smaller eddies and to those perpendicular modes. As the heater is turned off, the energy contained in the parallel modes would continue to feed to the perpendicular modes until finally all the modes are populated such that a Kolmogorov spectrum can be achieved. The build up of the perpendicular modes is likely to take longer time than is indicated by the solution of the Leith equation. Based on this idea, Joe started to develop a revised model of the bump, taking into account the spectrum anisotropy. But unfortunately, his physical conditions deteriorated rapidly starting from 2021, and he was not able to finish this work before he passed away in 2022.

\section{Summary and ongoing work}\label{Sec4}
We have performed a Leith-equation analysis on the time evolution of the turbulence energy spectrum with an initial form that is consistent with what we observed in our decaying He II counterflow experiment. Our results suggest that a key assumption made in the theoretical model developed by Joe Vinen and colleagues for explaining the bump puzzle is likely inadequate. According to Joe, this issue may be resolvable by taking into account the anisotropy of the initial energy spectrum.

To carry forward Joe's idea and research efforts, we recently started a numerical simulation using a revised Leith-diffusion model designed for homogeneous but anisotropic turbulence~\cite{Rubinstein-2017-CF}. Besides the Leith-equation simulation, we have also developed a stereoscopic molecular tagging technique to measure the normal-fluid velocities both parallel and perpendicular to the heat flux direction~\cite{Bao-2020-IJHMT}. Our preliminary data suggest that the energy spectrum of the parallel velocity component is much larger than that of the perpendicular velocity component, and this difference increases with decreasing $k$. Note that in the Hall-Vinen-Bekarevich-Khalatnikov (HVBK) model simulations conducted by Biferale \emph{et al.}, an isotropic driving force acting at large length scales was adopted to generate the turbulence~\cite{Biferale-2019-PRL,Biferale-2019-PRB}. This forcing scheme leads to comparable spectrum heights at small $k$ for both velocity components, which is unable to account for our observation. On the other hand, our earlier experimental and numerical studies revealed that the vortex-line density fluctuations in steady counterflow can lead to velocity fluctuations primarily in the heat flux direction~\cite{Mastracci-2018-PRF,Mastracci-2019-PRF-2,Yui-2020-PRL}. But these anitropic disturbances occur at relatively small length scales and are not expected to cause large-scale anisotropic flows. Nonetheless, Polanco and Krstulovic recently conducted HVBK-model simulations assuming a random isotropic driving force that acts only at small length scales~\cite{Polanco-2020-PRL}. They showed strikingly that the turbulence energy can inversely cascade to large length scales. We tried to repeat this HVBK-model simulation using a driving force at small length scales that fluctuates in both space and time but could not reproduce the reported inverse energy cascade. Through communications with these authors, we realized that the driving force adopted in their simulations was random in space but constant in time, which is probably not realistic in real thermal counterflow. We are currently devising new forms of anisotropic driving force in the HVBK model with the hope to better reproduce our experimental observations. These experimental and numerical results will be compiled and reported in a future publication.

\section*{Dedication to Joe Vinen}
The author W.G. first met Joe Vinen in 2008 at a workshop in San Antonio on low-temperature flow visualization, organized by Steven W. Van Sciver. They started collaborating since then. After W.G. joined Florida State University, Joe regularly visited W.G.'s lab every year to participate data analysis and result interpretation. Practically, Joe served the mentor role and motivated W.G. to investigate various aspects of quantum turbulence. Over the years, they jointly published over a dozen papers. Joe's passion for research and his kind guidance to younger generations will be remembered for years to come.

\section*{Acknowledgments}
The authors acknowledge the support provided by the National Science Foundation under Grant No. DMR-2100790 and the Gordon and Betty Moore Foundation through Grant GBMF11567. The work was done at the National High Magnetic Field Laboratory at Florida State University, which is supported by the National Science Foundation Cooperative Agreement No. DMR-1644779 and the state of Florida.

\bibliography{qtbib}
\end{document}